\documentclass[12pt]{article}
\usepackage{cite}
\usepackage{amsmath, graphicx, blkarray,pdflscape}
\usepackage{tabularx, makecell,amsfonts}
\usepackage{xcolor,amsthm,geometry,rotating, booktabs,mathtools}
\usepackage{float}
\usepackage{pgf,tikz}
\usepackage{multirow}
\usepackage{natbib}
\usetikzlibrary{shapes,decorations,arrows,calc,arrows.meta,fit,positioning}
\tikzset{
	-Latex,auto,node distance =1 cm and 1 cm,semithick,
	state/.style ={ellipse, draw, minimum width = 0.7 cm},
	point/.style = {circle, draw, inner sep=0.04cm,fill,node contents={}},
	bidirected/.style={Latex-Latex,dashed},
	el/.style = {inner sep=2pt, align=left, sloped}
}

\DeclareMathOperator*{\argmax}{arg\,max}

\newcommand{\blind}{0}
\usepackage{natbib}
\setcitestyle{authoryear,open={(},close={)}} 

\begin{document}
	\def\spacingset#1{\renewcommand{\baselinestretch}%
		{#1}\small\normalsize} \spacingset{1}
	
	\if0\blind
	{
		\title{\bf Statistical methods for clustered competing risk data when the event types are only available in a training dataset}
		\author{Yujie Wu\\
			\textit{Department of Biostatistics, Harvard University}\\
			and \\
			Molin Wang \\
			\textit{Departments of Biostatistics and Epidemiology, Harvard University,}\\
			\textit{Channing Division of Network Medicine, Brigham and Women's Hospital,}\\\textit{and Harvard Medical School}}
		\date{}
		\maketitle
	} \fi
	
	\begin{abstract}
		We develop methods to analyze clustered competing risks data when the event types are only available in a training dataset and are missing in the main study. We propose to estimate the exposure effects through the cause-specific proportional hazards frailty model where random effects are introduced into the model to account for the within-cluster correlation. We propose a weighted penalized partial likelihood method where the weights represent the probabilities of the occurrence of events, and the weights can be obtained by fitting a classification model for the event types on the training dataset. Alternatively, we propose an imputation approach where the missing event types are imputed based on the predictions from the classification model. We derive the analytical variances, and evaluate the finite sample properties of our methods in an extensive simulation study. As an illustrative example, we apply our methods to estimate the associations between tinnitus and metabolic, sensory and metabolic+sensory hearing loss in the Conservation of Hearing Study Audiology Assessment Arm.
	\end{abstract}
	
 	\newpage
\newpage
\spacingset{1.9} 
	
	\section{Introduction}
In epidemiological and clinical studies, it is common that participants are at risk of more than one competing events. For example, in our motivating hearing loss epidemiological study, the following three types of age-related hearing loss were observed: metabolic, sensory and metabolic+sensory. Only one out of these three types can occur for an individual with hearing loss; thus, the events/failure types can be treated as competing risks \citep{gupta2019type, lin2020cigarette}. Furthermore, the event type data may not always be available for participants who have experienced the failure. In our motivating study, the hearing loss types for both ears were determined by hearing experts and they are available only for participants in an training dataset.  Also, the two ears' event type data within a participant are correlated, and accounting for correlation between clustered outcomes is essential for valid statistical inference. In this paper, we propose statistical methods to analyze clustered competing risks data when the types of the competing events are only available for participants in a training dataset.
		
When the types of failure events are observed, the competing risks data is commonly modelled through the cause-specific proportional hazards model \citep{prentice1978analysis}. To account for the clustering, we use frailty models with random effects included in the cause-specific proportional hazards model, taking into account the within-cluster correlations of the cause-specific hazards \citep{hougaard1995frailty, dharmarajan2018evaluating, wang2020penalized}.

	
	Several methods have been proposed to address the issue of missing event types for the traditional cause-specific models when the survival outcomes are independent. Goetghebeur and Ryan (1995) proposed method to estimate the regression parameters assuming the baseline hazards of different causes are proportional \citep{goetghebeur1995analysis}. \cite{lu2001multiple} developed a multiple imputation approach to impute missing cause of failures when fitting the cause-specific proportional hazards models. \cite{craiu2004inference} used an EM algorithm to estimate the parameters, and \cite{hyun2012proportional} proposed an augmented inverse probability weighted estimator for the regression parameters that have the double-robustness property. However, there is no existing method addressing the missing event type issue for the cause-specific proportional hazards frailty models in the clustered survival data settings.

	Our work is motivated by the Conservation of Hearing Study (CHEARS) that evaluates risk factors for hearing loss in Nurses Health Study II (NHS II), an ongoing cohort study containing 116,430 female registered nurses in the United States \citep{curhan2018adherence}. In CHEARS, participants' hearing statuses and, for those who reported hearing loss, the age at the occurrence of the incident hearing loss, were collected through questionnaires. The CHEARS Audiology Assessment Arm (AAA) is a sub-cohort of CHEARS where 3,749 participants' audiometric hearing thresholds were measured by audiologists. However, the hearing loss types are missing in AAA. In an external training data set from the Medical University of South Carolina (MUSC) database, both the audiometric hearing thresholds data and the phenotype information are available among 338 ears \citep{dubno2013classifying}. 

	In this paper, we propose methods to address the statistical challenge due to unknown event types in cause-specific proportional hazards frailty models. The paper is organized as follows. In Section 2, we propose a weighted estimating equations approach and an imputed estimating equations approach. Our methods can be used to handle both scenarios when the training dataset is either a subset of the main study or an external data set. In Section 3, we evaluate the performance of our proposed methods through extensive simulation studies. In Section 4, we apply the methods to analyze the associations between tinnitus and metabolic, sensory and metabolic+sensory hearing loss in AAA, and Section 5 concludes the paper.

	\section{Method}
	\subsection{Estimating equations when event types are available}
	Let $i (i=1,\ldots, N)$ index clusters (participants in our motivating example), $j (j=1,\ldots,n_i)$ the units (ear in our example) within a cluster, and $k (k=1,\ldots,K)$ the competing events (hearing loss types in our example). For unit $j$ in the $i$-th cluster, let $T^e_{ij}$ and $C_{ij}$ denote the event time and censoring time, respectively, and we observe $T_{ij}=\min(T^e_{ij}, C_{ij} )$. The event indicator $\delta_{ij}\in\{0,\ldots, K\}$ denotes the type of the event and we let $\delta_{ij}=0$ if $T^e_{ij}>C_{ij}$. Moreover, let $\Delta_{ij}=I(T^e_{ij}\le C_{ij})$ be the censoring indicator; $\Delta_{ij}=0$ if participant is censored at the end of follow-up, and $\Delta_{ij}=1$ otherwise. Note that, $\Delta_{ij}$ is always observed even though the specific event types $\delta_{ij}$ are missing for participants who have experienced the events before the censoring time. The cause-specific hazard function follows the proportional hazards model:
\begin{equation*}
	\lambda_{ijk}(t)=\lambda_{0k}(t)\exp\left(\boldsymbol{X}_{ij}^T\boldsymbol{\beta}_k+{v_{ik}}\right), k=1,\ldots, K,
\end{equation*}
where $\boldsymbol{X}_{ij}$ and $\boldsymbol{\beta}_k$ are both column vectors containing the covariates of interest and the corresponding cause-specific regression coefficients, respectively, $\lambda_{0k}(t)$ is the cause-specific baseline hazard function, and $v_{ik}$ is the cluster-and event-specific random effect that accounts for the correlation of cause-specific hazards between units within the same cluster. Let $\boldsymbol{v}_i=[v_{i1}, v_{i2}, \ldots, v_{iK}]^T$ be a column vector of the random effects for cluster $i$, and we assume $\boldsymbol{v}_i$ follows a multivariate normal distribution with mean $\boldsymbol{0}$ and variance-covariance matrix $\boldsymbol{D}_i(\boldsymbol{\theta})$, where $\boldsymbol{\theta}$ is the unknown parameters in the variance-covariance matrix. 

When the event indicator $\delta_{ij}$ is available, the likelihood contribution from cluster $i$ conditional on the random effects $\boldsymbol{v}_{i}$ and covariates is:
\begin{equation*}
	\begin{split}
		L_{i}(\boldsymbol{\beta},\lambda_{01}(\cdot),\ldots,\lambda_{0K}(\cdot)|\boldsymbol{v}_{i}, \boldsymbol{X}_{i1,},\ldots,\boldsymbol{X}_{i,n_i})=&\prod_{j=1}^{n_i}\prod_{k=1}^{K}\Bigg\{\left(\lambda_{0k}(T_{ij})\exp(\boldsymbol{X}_{ij}^T\boldsymbol{\beta}_k+v_{ik})\right)^{I(\delta_{ij}=k)}\\
		&\times\exp\left[-\Lambda_{0k}(T_{ij})\exp(\boldsymbol{X}_{ij}^T\boldsymbol{\beta}_k+v_{ik})\right]\Bigg\},
	\end{split}
\end{equation*}
where $\Lambda_{0k}(t)$ is the cumulative baseline hazard function for the $k$-th competing event. The marginal likelihood of the whole data set can be obtained by integrating out the random effects as follows.
\begin{equation}
	\begin{split}
		L_M(\boldsymbol{\beta},\lambda_{01}(\cdot),\ldots,\lambda_{0K}(\cdot)| \boldsymbol{X}_{1},\ldots,\boldsymbol{X}_N)=&\int\prod_{i=1}^{N}\Bigg\{\Big(\prod_{j=1}^{n_i}\prod_{k=1}^{K}\left[\lambda_{0k}(T_{ij})\exp(\boldsymbol{X}_{ij}^T\boldsymbol{\beta}_k+v_{ik})\right]^{I(\delta_{ij}=k)}\\
		&\times\exp\left[-\Lambda_{0k}(T_{ij})\exp(\boldsymbol{X}_{ij}^T\boldsymbol{\beta}_k+v_{ik})\right]\Big)\times p(\boldsymbol{v}_i;\boldsymbol{D}_i(\boldsymbol{\theta}))\Bigg\}\,d\boldsymbol{v}_i,
	\end{split}
	\label{marginal_like_full}
\end{equation}
where $\boldsymbol{X}_i=[\boldsymbol{X}_{i1}^T,\ldots, \boldsymbol{X}_{i,n_i}^T]^T$, and $p(\boldsymbol{v}_i;\boldsymbol{D}_i(\boldsymbol{\theta}))$ is the probability density function for $\boldsymbol{v}_i$. We further assume the random effects between different clusters are independent, and therefore the collection of random effects from all clusters $\boldsymbol{v}=[\boldsymbol{v}_1^T,\ldots,\boldsymbol{v}_N^T]^T$ jointly follows a multivariate normal distribution with mean $\boldsymbol{0}$ and variance-covariance matrix $\boldsymbol{D}(\boldsymbol{\theta})$, where $\boldsymbol{D}(\boldsymbol{\theta})$ is a block diagonal matrix, with the $i$-th diagonal block being $\boldsymbol{D}_i(\boldsymbol{\theta})$. 
 A Laplace approximation can be applied to the marginal likelihood, leading to the following approximate marginal log-likelihood \citep{wang2020penalized}:
\begin{equation}
	\ell_M=\log(L_M)\approx -\frac{1}{2}\log|\boldsymbol{D}(\boldsymbol{\theta})| - \boldsymbol{K}(\widetilde{\boldsymbol{\boldsymbol{v}}})-\frac{1}{2}\log|\boldsymbol{K}_2(\widetilde{\boldsymbol{\boldsymbol{v}}})|,
	\label{approx_log_like}
\end{equation}
where $\boldsymbol{K}(\widetilde{\boldsymbol{\boldsymbol{v}}})$ and $\boldsymbol{K}_2(\widetilde{\boldsymbol{\boldsymbol{v}}})$ are defined in the supplementary material. Following the arguments in \cite{ripatti2000estimation}, \cite{wang2020penalized}, and \cite{dharmarajan2018evaluating}, the regression parameters can be estimated by maximizing the following Penalized Partial Likelihood (PPLL):
	\begin{equation*}
	\scriptsize
	PPLL=	 {\color{black}{-\frac{1}{2}\boldsymbol{v}^T\boldsymbol{D}(\boldsymbol{\theta})^{-1}\boldsymbol{v}}}+\sum_{i=1}^{N}\sum_{j=1}^{n_i}\sum_{k=1}^{K}\Bigg\{I(\delta_{ij}=k)\left[\boldsymbol{X}_{ij}^T\boldsymbol{\beta}_k+v_{ik}-\log\sum_{i'=1}^{N}\sum_{j'=1}^{n_i}Y_{i'j'}(T_{ij})\exp\left( \boldsymbol{X}_{i'j'}\boldsymbol{\beta}_k^T+v_{i'k} \right)\   \right]  \Bigg\},
\end{equation*}
where $-\frac{1}{2}\boldsymbol{v}^T\boldsymbol{D}(\boldsymbol{\theta})^{-1}\boldsymbol{v}$ can be regarded as the penalty term applied to the usual partial likelihood for the Cox model. See the supplementary material for a derivation of the PPLL.

The parameters can be estimated through an iterative approach. For a given $\boldsymbol{\theta}$, $\boldsymbol{\beta}$ can be estimated by maximizing the PPLL. Therefore, the estimating function for $\boldsymbol{\beta}$ is:
\begin{equation}
	\footnotesize
	\boldsymbol{U}(\boldsymbol{\beta}_k)=\frac{\partial PPLL}{\partial \boldsymbol{\beta}_k}=\sum_{i=1}^{N}\sum_{j=1}^{n_i}\Bigg\{I(\delta_{ij}=k)\left[\boldsymbol{X}_{ij} - \frac{\sum_{i'=1}^{N}\sum_{j'=1}^{n_i}Y_{i'j'}(T_{ij})\exp\left( \boldsymbol{X}_{i'j'}\boldsymbol{\beta}_k^T+{v}_{i'k} \right)\boldsymbol{X}_{i'j'}}{\sum_{i'=1}^{N}\sum_{j'=1}^{n_i}Y_{i'j'}(T_{ij})\exp\left( \boldsymbol{X}_{i'j'}\boldsymbol{\beta}_k^T+{v}_{i'k} \right)} \right]\Bigg\},
	\label{score_beta}
\end{equation}
$k=1\ldots, K$, and $\boldsymbol{\boldsymbol{v}}$ is the solution to the following estimating function:


\begin{equation}
	\footnotesize
	\boldsymbol{U}(\boldsymbol{v})=\frac{\partial PPLL}{\partial \boldsymbol{v}}=\sum_{i=1}^{N}\sum_{j=1}^{n_i}\sum_{k=1}^{K}\Bigg\{I(\delta_{ij}=k)\left[\boldsymbol{R}_{ik} - \frac{\sum_{i'=1}^{N}\sum_{j'=1}^{n_i}Y_{i'j'}(T_{ij})\exp\left( \boldsymbol{X}_{i'j'}\boldsymbol{\beta}_k^T+{v}_{i'k} \right)\boldsymbol{R}_{i'k}}{\sum_{i'=1}^{N}\sum_{j'=1}^{n_i}Y_{i'j'}(T_{ij})\exp\left( \boldsymbol{X}_{i'j'}\boldsymbol{\beta}_k^T+{v}_{i'k} \right)} \right]\Bigg\}-\boldsymbol{D}(\boldsymbol{\theta})^{-1}\boldsymbol{v}.
	\label{score_v},
\end{equation}
where $\boldsymbol{R}_{ik}$ is a $NK\times 1$ column vector containing 0 and 1's, with the $K(i-1)+k$-th element being 1 and the rest elements being 0.

The variance components for the random effects can be estimated by maximizing the marginal likelihood in Equation (\ref{approx_log_like}), given the current estimate of $\boldsymbol{\beta}$ and $\boldsymbol{v}$. Taking derivative of the PPLL with respect to $\boldsymbol{\theta}$, we have the following estimating function for $\boldsymbol{\theta}$:
\begin{equation}
	\footnotesize
	{\boldsymbol{U}}(\boldsymbol{\theta})=-\frac{1}{2}\left[   tr\left( \boldsymbol{D}^{-1}\frac{\partial\boldsymbol{D}}{\partial\boldsymbol{\theta}} \right) -\boldsymbol{v}^T\boldsymbol{D}^{-1}\frac{\partial\boldsymbol{D}}{\partial\boldsymbol{\theta}}\boldsymbol{D}^{-1}\boldsymbol{v}+tr\left( \boldsymbol{K}_{2, PPLL}(\boldsymbol{v})^{-1}  \frac{\partial\boldsymbol{D}}{\partial\boldsymbol{\theta}}\right)\right],
	\label{score_theta}
\end{equation}
where $\boldsymbol{K}_{2, PPLL}=\frac{\partial^2 {PPLL}}{\partial\boldsymbol{v}\partial\boldsymbol{v}^T}$.

\subsection{Weighted estimating equations with unknown event types}\label{weighted}

In this section, we develop a weighted estimating equations approach when the event types are unknown. Note that even though the event type $\delta_{ij}$ is not available, the censoring indicator $\Delta_{ij}$ is always observed. In this paper we focus on the setting where a collection of variables $\boldsymbol{W}_{ij}$ that are predictive of the event types are observed; this setting is common in epidemiological studies. In the motivating example, $\boldsymbol{W}_{ij}$'s are the hearing thresholds which are used to classify hearing loss phenotypes.  When the true event types $\delta_{ij}$ is unknown, we propose to use the following weighted estimating equations in place of Equation (\ref{score_beta}), and detailed derivations are given in supplementary material section 1.2:
		\begin{equation}
			\scriptsize
			\widetilde{\boldsymbol{U}}(\boldsymbol{\beta}_k)=\boldsymbol{E}(\boldsymbol{U}(\boldsymbol{\beta}_k)|\boldsymbol{W}, \boldsymbol{\Delta})=\sum_{i=1}^{N}\sum_{j=1}^{n_i}\Bigg\{P(\delta_{ij}=k|\boldsymbol{W}_{ij}, \Delta_{ij}=1)\left[\boldsymbol{X}_{ij} - \frac{\sum_{i'=1}^{N}\sum_{j'=1}^{n_i}Y_{i'j'}(T_{ij})\exp\left( \boldsymbol{X}_{i'j'}\boldsymbol{\beta}_k^T+{v}_{i'k} \right)\boldsymbol{X}_{i'j'}}{\sum_{i'=1}^{N}\sum_{j'=1}^{n_i}Y_{i'j'}(T_{ij})\exp\left( \boldsymbol{X}_{i'j'}\boldsymbol{\beta}_k^T+{v}_{i'k} \right)} \right]\Bigg\}=\boldsymbol{0}.
			\label{score_beta2}
		\end{equation}
		Here, the event type indicator $I(\delta_{ij}=k)$ in Equation (\ref{score_beta}) is replaced with the corresponding conditional probability of the occurrence of the $k$-th event given $\boldsymbol{W}_{ij}$ and $\Delta_{ij}=1$: $P(\delta_{ij}=k|\boldsymbol{W}_{ij}, \Delta_{ij}=1)$. For presentational simplicity, we use $P(\delta_{ij}=k)$ to denote it thereafter. Note that the estimating equation (\ref{score_beta2}) corresponds to maximizing the following weighted PPLL:
		\begin{equation*}
			\begin{split}
				\widetilde{PPLL}&=	-\frac{1}{2}\boldsymbol{v}^T\boldsymbol{D}(\boldsymbol{\theta})^{-1}\boldsymbol{v}\\
				&+\sum_{i=1}^{N}\sum_{j=1}^{n_i}\sum_{k=1}^{K}\Bigg\{P(\delta_{ij}=k)\left[\boldsymbol{X}_{ij}^T\boldsymbol{\beta}_k+v_{ik}-\log\sum_{i'=1}^{N}\sum_{j'=1}^{n_i}Y_{i'j'}(T_{ij})\exp\left( \boldsymbol{X}_{i'j'}\boldsymbol{\beta}_k^T+v_{i'k} \right)\   \right]  \Bigg\}.\\
			\end{split}
\end{equation*}

Similarly, the estimating equations for $\boldsymbol{v}$ and $\boldsymbol{\theta}$ can be obtained by make the same replacement on the indicator function $I(\delta_{ij}=k)$ in Equations (\ref{score_v}) and (\ref{score_theta}):
\begin{equation*}
	\footnotesize
	\widetilde{\boldsymbol{U}}(\boldsymbol{v})=\sum_{i=1}^{N}\sum_{j=1}^{n_i}\sum_{k=1}^{K}\Bigg\{P(\delta_{ij}=k)\left[\boldsymbol{R}_{ik} - \frac{\sum_{i'=1}^{N}\sum_{j'=1}^{n_i}Y_{i'j'}(T_{ij})\exp\left( \boldsymbol{X}_{i'j'}\boldsymbol{\beta}_k^T+{v}_{i'k} \right)\boldsymbol{R}_{i'k}}{\sum_{i'=1}^{N}\sum_{j'=1}^{n_i}Y_{i'j'}(T_{ij})\exp\left( \boldsymbol{X}_{i'j'}\boldsymbol{\beta}_k^T+{v}_{i'k} \right)} \right]\Bigg\}-\boldsymbol{D}(\boldsymbol{\theta})^{-1}\boldsymbol{v}=\boldsymbol{0},
\end{equation*}

\begin{equation*}
	\footnotesize
	\widetilde{\boldsymbol{U}}(\boldsymbol{\theta})=-\frac{1}{2}\left[   tr\left( \boldsymbol{D}^{-1}\frac{\partial\boldsymbol{D}}{\partial\boldsymbol{\theta}} \right) -\boldsymbol{v}^T\boldsymbol{D}^{-1}\frac{\partial\boldsymbol{D}}{\partial\boldsymbol{\theta}}\boldsymbol{D}^{-1}\boldsymbol{v}+tr\left( \widetilde{\boldsymbol{K}}_{2, PPLL}(\boldsymbol{v})^{-1}  \frac{\partial\boldsymbol{D}}{\partial\boldsymbol{\theta}}\right)\right]=\boldsymbol{0},
\end{equation*}
where $\widetilde{\boldsymbol{K}}_{2, PPLL}=\frac{\partial^2 \widetilde{PPLL}}{\partial\boldsymbol{v}\partial\boldsymbol{v}^T}$.

The probability $P(\delta_{ij}=k)$ in the estimating equations can be estimated by fitting a classification model in a subset of the participants or an external study where the event types and $\boldsymbol{W}_{ij}$ that are predictive of the event types are available. For example, in MUSC, both the hearing loss phenotypes and audiometric hearing thresholds are available; therefore a classification model in MUSC for the three phenotypes can be fitted based on the audiometric threshold measurements, and the model can be applied to obtain  estimates of $P(\delta_{ij}=k)$ for participants in AAA. Some common classification models that can be chosen includes quadratic discriminant analysis, support vector machine, random forests etc. Note that the classification model is fitted among participants that have experienced the events (i.e. $\Delta_{ij}=1$).

Based on the sandwich variance approach and ignoring the fact that the probabilities $P(\delta_{ij}=k)$ are estimated, the variance of $(\widehat{\boldsymbol{\beta}}^T, \widehat{\boldsymbol{v}}^T)^T$ can be estimated by \citep{gray1992flexible, dharmarajan2018evaluating}:
\begin{equation}
	\widehat{\boldsymbol{V}}(\widehat{\boldsymbol{\beta}},\widehat{\boldsymbol{v}})=\boldsymbol{
	H}(\widehat{\boldsymbol{\beta}},\widehat{\boldsymbol{v}})^{-1}\boldsymbol{\mathcal{I}}(\widehat{\boldsymbol{\beta}},\widehat{\boldsymbol{v}})\boldsymbol{
	H}(\widehat{\boldsymbol{\beta}},\widehat{\boldsymbol{v}})^{-1},
		\label{variance}
\end{equation}
where
\begin{equation*}
	\boldsymbol{H}(\boldsymbol{\beta},\boldsymbol{v})=\boldsymbol{\mathcal{I}}(\boldsymbol{\beta},\boldsymbol{v})+\begin{bmatrix}
		\boldsymbol{0}&\boldsymbol{0}\\
		\boldsymbol{0}&\boldsymbol{D}^{-1}
	\end{bmatrix},
\end{equation*}
 $\boldsymbol{\mathcal{I}}(\boldsymbol{\beta}, \boldsymbol{v})=\begin{bmatrix}
	\boldsymbol{\mathcal{I}}_{\boldsymbol{\beta}\boldsymbol{\beta}}&\boldsymbol{\mathcal{I}}_{\boldsymbol{\beta}\boldsymbol{v}}\\
	\boldsymbol{\mathcal{I}}_{\boldsymbol{\beta}\boldsymbol{v}}^T&\boldsymbol{\mathcal{I}}_{\boldsymbol{v}\boldsymbol{v}}
\end{bmatrix}$ is the usual information matrix of the weighted partial likelihood function without the penalty ($-\frac{1}{2}\boldsymbol{v}^T\boldsymbol{D}(\boldsymbol{\theta})^{-1}\boldsymbol{v}$), and the formula for the elements in the information matrix is given in Appendix.

{\color{black}{ As discussed in \cite{lok2021estimating}, this variance formula typically leads to conservative inferences. Another option is using the inverse of the Hessian matrix $\boldsymbol{
		H}(\widehat{\boldsymbol{\beta}},\widehat{\boldsymbol{v}})^{-1}$ as the variance estimator, and it may be slightly more conservative than the sandwich variance estimator \citep{therneau2003penalized}. In the Supplementary Material Section 2, we describe, through a toy example, how to perform data duplication and manipulation so that we can use the existing \texttt{R} package \texttt{coxme} to fit our proposed model. The inverse of Hessian matrix is used as the variance estimator in this package, and thus, we adopt $\boldsymbol{
		H}(\widehat{\boldsymbol{\beta}},\widehat{\boldsymbol{v}})^{-1}$ as the variance estimator in our simulation studies and real data application.
	
	}}

\subsection{Imputed estimating equations with unknown event type}
In section \ref{weighted}, the classification model is used to obtain the predicted event probabilities $P(\delta_{ij}=k)$ for individuals in the main study. Alternatively, the event types can directly be imputed based on the estimated event probabilities, where we assign the event type to be $k$, if the $k$-th event has the highest estimated probability:
\begin{equation*}
	\widehat{\delta}_{ij}=\argmax_k \widehat{P}(\delta_{ij}=k), \text{ if } \Delta_{ij}=1.
 \end{equation*}

Note that the imputation should not be performed for participants that are censored. Therefore, the regression coefficients can be estimated through solving Equations (\ref{score_beta})-(\ref{score_theta}) by replacing $I(\delta_{ij}=k)$ with $I(\widehat{\delta}_{ij}=k)$. Inference of the estimated parameters can be performed using the sandwich variance estimator similar to Equation (\ref{variance}), except that we replace $\widehat{P}(\delta_{ij}=k)$ with $I(\widehat{\delta}_{ij}=k)$. The parameters can also be estimated using the \texttt{coxme} function without the weights, and the slightly more conservative inverse Hessian matrix $\boldsymbol{
	H}(\widehat{\boldsymbol{\beta}},\widehat{\boldsymbol{v}})^{-1}$ is available from the package as the variance estimator.

\section{Simulation}

In this section, we conduct a simulation study to evaluate the performance of our proposed methods. We generate $N=500, 1000$ clusters with $n_i=2$ units within each cluster in the main study, where the true event types are unobserved. We set $K=2$ competing events with the following cause-specific proportional hazard functions:
\begin{equation*}
	\lambda_{ijk}(t)=\lambda_{0k}(t)\exp\left(X_{ij}\beta_k+v_{ik}\right), k=1,2,
\end{equation*}
where we fix the baseline hazards to be constant: $\lambda_{0k}(t)=1$ for $k=1,2$. The one-dimensional covariate $X_{ij}$ is generated from a Bernoulli distribution with success probability 0.5, and the random effects within a cluster $i$ are generated from a multivariate normal distribution: $\begin{bmatrix}
	v_{i1}\\
	v_{i2}
\end{bmatrix}\sim\text{MVN}\left(\boldsymbol{0}, \begin{bmatrix}
0.1& 0.1\rho\\
0.1\rho&0.1
\end{bmatrix}\right)$, and we let $\rho=0.2, 0.5$ for different degrees of correlation. The regression parameters $(\beta_1, \beta_2)$ are chosen among $\{(\log(1.25), \log(1.5)), (\log(1.5), \log(1.75)), (\log(1.5), \log(1.75))\}$.

The true event types are determined by an event predictor ${W}_{ij}$ (e.g. the audiometric hearing thresholds in hearing loss study). For units $j=1, 2$ within a cluster we generate $W_{i1}$ and $W_{i2}$ from a bivariate normal distribution: $\begin{bmatrix}
	W_{i1}\\
	W_{i2}
\end{bmatrix}\sim \text{MVN}\left(\begin{bmatrix}
	\sum_{k=0}^2\mu_{k}I(\delta_{i1}=k)\\
	\sum_{k=0}^2\mu_{k}I(\delta_{i2}=k),
\end{bmatrix},     	\begin{bmatrix}
	1& 0.25\\
	0.25&1
\end{bmatrix}\right)$, where $\mu_0$ is the mean value of $W_{ij}$ if the participant is censored without experiencing any event, while $\mu_1$ and $\mu_2$ represent the mean values of $W_{ij}$ if the participant experiences the first or the second event. When the differences among $\mu_k, k=0,1,2$, are larger, $W_{ij}$ can be used to distinguish between different event types with more accuracy. In our simulation study, we set the difference $\gamma\in\{2.5, 3, 3.5\}$, where $\gamma=\mu_1-\mu_0=\mu_2-\mu_1$. In a training dataset, both the event types and $W_{ij}$ are available, and we use the multinomial logistic regression to build the classification model, which is then used to obtain the event probabilities for participants in the main study if using the weighted estimating equations approach or to impute the event types if using the imputed estimating equations approach. We set the sample size of training dataset to be 50 or 100.

Shown in Table \ref{sim1} and supplementary table 4 are the simulation results when the correlation coefficient between the random effects are set to be 0.5 and 0.2, respectively.  We observe that both the weighted and imputed estimating equation estimators have percent biases that are generally smaller than 5\%, and the coverage rates of the 95\% confidence intervals (CI) based on the inverse of the Hessian matrix are close to the 95\% nominal level. Increasing the sample size of the main study or the training dataset lead to decreased percent biases of the estimated regression coefficients. Moreover, when the event predictor $W_{ij}$'s are well separated (i.e. $\gamma$ is large), the regression coefficients can be estimated with less biases.

The proposed weighted and imputed estimating equations approaches have comparable finite sample performances. The imputed estimating equation estimators have less percent biases when the covariate effects are large (i.e. $\beta_1=\log(1.5), \beta_2=\log(1.75)$), while the weighted estimating equations approach is typically more efficient as the empirical standard errors of the parameter estimates over the 1000 simulation replicates are in general smaller than those from the imputed estimating equations approach.

\begin{table}[htb]
	\centering
	\tiny
	\caption{Simulation study results when $\rho=0.5$ for the correlation coefficient of the random effects. The sample size of the main study is chosen among 500 and 1000, and the sample size of the training study for fitting the event predicting model is set to be 50 or 100. A total of 1000 simulation replicates are conducted, and the percent biases (\%bias) of the coefficient estimates, empirical standard errors (ESE) and coverage probabilities of the 95\% confidence intervals are reported for both the weighted estimating equations and imputed estimating equations approach.}
	\begin{tabular}{cccccccc}
		\hline\hline
	\multirow{2}{*}{Sample size }& 	\multirow{2}{*}{Method}	 &\multirow{2}{*}{Parameters} &	\multirow{2}{*}{$\gamma$}	& \multicolumn{2}{ c }{Percent bias (\%)(ESE)}  & \multicolumn{2}{ c }{Coverage Probability}  \\
 	&&&&$\beta_1$&$\beta_2$&$\beta_1$&$\beta_2$\\
		\hline
 \multirow{12}{*}{$(1000,50)$}& \multirow{6}{*}{Weighted}&\multirow{3}{*}{$ (\log(1.5), \log(1.5)) $}	 & 2.5   & -0.1\% (0.069) & 0.1\% (0.063) & 0.95  & 0.97 \\    
 & &  & 3     & -0.4\% (0.068) & 0.2\% (0.068) & 0.95  & 0.96 \\    
& &  & 3.5   & -0.2\% (0.065) & -1.1\% (0.066) & 0.95  & 0.96 \\   \cmidrule{3-8}
 &&\multirow{3}{*}{$ (\log(1.5), \log(1.75)) $}& 2.5   & 6.0\% (0.069) & -4.9\% (0.066) & 0.94  & 0.92 \\    
& &  & 3     & 4.3\% (0.066) & -2.8\% (0.065) & 0.96  & 0.96 \\    
& &  & 3.5   & 2.4\% (0.072) & -2.1\% (0.067) & 0.94  & 0.95 \\    \cmidrule{2-8}

& \multirow{6}{*}{Imputed}&\multirow{3}{*}{$ (\log(1.5), \log(1.5)) $}& 2.5   & -1.0\% (0.071) & -0.7\% (0.069) & 0.95  & 0.95 \\   
 & &  & 3     & -0.1\% (0.070) & -1.3\% (0.071) & 0.95  & 0.95 \\   
& & & 3.5   & 0.2\% (0.068) & -1.0\% (0.070) & 0.95  & 0.95 \\   \cmidrule{3-8}
& & \multirow{3}{*}{$ (\log(1.5), \log(1.75)) $} & 2.5   & 3.9\% (0.072) & -3.4\% (0.072) & 0.94  & 0.93 \\    &&& 3     & 2.3\% (0.073) & -1.6\% (0.071) & 0.94  & 0.94 \\    
&& & 3.5   & 0.3\% (0.071) & -1.1\% (0.072) & 0.95  & 0.93 \\  \hline

 \multirow{12}{*}{$(1000,100)$}& \multirow{6}{*}{Weighted}&\multirow{3}{*}{$ (\log(1.5), \log(1.5)) $}	& 2.5   & -0.7\% (0.064) & -0.4\% (0.065) & 0.96  & 0.96 \\          
&       &       & 3     & -0.7\% (0.066) & -0.3\% (0.067) & 0.96  & 0.96 \\         
 &       &       & 3.5   & -0.9\% (0.068) & -0.0\% (0.069) & 0.94  & 0.94 \\ \cmidrule{3-8}        
  &       &   \multirow{3}{*}{$ (\log(1.5), \log(1.75)) $}    & 2.5   & 5.6\% (0.066) & -4.7\% (0.065) & 0.96  & 0.94 \\          
  &       &       & 3     & 3.5\% (0.068) & -3.3\% (0.067) & 0.95  & 0.94 \\          &       &       & 3.5   & 2.0\% (0.070) & -2.2\% (0.068) & 0.94  & 0.95 \\    \cmidrule{2-8}

& \multirow{6}{*}{Imputed}&\multirow{3}{*}{$ (\log(1.5), \log(1.5)) $}	 & 2.5   & -1.1\% (0.070) & -0.1\% (0.072) & 0.95  & 0.93 \\          
   &       &       & 3     & -1.1\% (0.069) & -0.6\% (0.071) & 0.95  & 0.93 \\         
    &       &       & 3.5   & -0.7\% (0.068) & -1.2\% (0.069) & 0.95  & 0.94 \\       \cmidrule{3-8}   
 &&\multirow{3}{*}{$ (\log(1.5), \log(1.75)) $}& 2.5   & 2.9\% (0.073) & -3.2\% (0.072) & 0.94  & 0.94 \\          
    &       &       & 3     & 2.4\% (0.072) & -2.3\% (0.071) & 0.94  & 0.93 \\          
    &       &       & 3.5   & 1.3\% (0.073) & -1.5\% (0.069) & 0.94  & 0.94 \\  \hline
    
 \multirow{12}{*}{$(500,50)$}& \multirow{6}{*}{Weighted}&\multirow{3}{*}{$ (\log(1.5), \log(1.5)) $}& 2.5   & 0.8\% (0.094) & 0.9\% (0.090) & 0.96  & 0.96 \\          
     &       &       & 3     & 0.2\% (0.094) & -1.3\% (0.093) & 0.95  & 0.96 \\       
        &       &       & 3.5   & -0.6\% (0.093) & -0.7\% (0.100) & 0.95  & 0.94 \\       \cmidrule{3-8}        
           &       &    \multirow{3}{*}{$ (\log(1.5), \log(1.75)) $}   & 2.5   & 6.3\% (0.093) & -3.9\% (0.092) & 0.95  & 0.95 \\         
            &       &       & 3     & 4.0\% (0.094) & -3.8\% (0.093) & 0.96  & 0.95 \\         
             &       &       & 3.5   & 3.2\% (0.098) & -1.7\% (0.091) & 0.95  & 0.96 \\  \cmidrule{2-8}

& \multirow{6}{*}{Imputed}&\multirow{3}{*}{$ (\log(1.5), \log(1.5)) $}	  & 2.5   & -0.2\% (0.098) & -0.4\% (0.098) & 0.95  & 0.95 \\          
&       &       & 3     & -0.8\% (0.099) & 0.4\% (0.096) & 0.94  & 0.95 \\          
&       &       & 3.5   & -1.0\% (0.098) & 0.9\% (0.100) & 0.95  & 0.95 \\    \cmidrule{3-8}               
&       &     \multirow{3}{*}{$ (\log(1.5), \log(1.75)) $}     & 2.5   & 3.0\% (0.104) & -2.8\% (0.100) & 0.94  & 0.94 \\          
&       &       & 3     & 2.1\% (0.095) & -2.6\% (0.100) & 0.96  & 0.95 \\          
&       &       & 3.5   & 0.2\% (0.100) & -1.7\% (0.098) & 0.94  & 0.95 \\

		\hline\hline
	\end{tabular}
\label{sim1}
\end{table}

\section{Real data}

In this section, we apply our proposed methods to investigate the associations between tinnitus and metabolic, sensory, and metabolic+sensory hearing losses in CHEARS AAA. Studies have shown that tinnitus is associated with elevated audiometric hearing thresholds, but there is limited work investigating the associations between tinnitus and different hearing loss phenotypes \citep{curhan2021tinnitus}. In the main study of the analysis, the CHEARS AAA study, the event time was collected based on participants' self-reports. If a participant reported moderate or more severe hearing loss, we consider them having one of the hearing loss phenotypes \citep{lin2020cigarette}. Since the phenotype information is missing in AAA, we use the MUSC database as the training dataset to build the classification model of the hearing loss phenotypes. In the MUSC database, out of 338 ears, 11\% were labelled as older-normal (i.e. without hearing loss) by hearing experts, 25\% as metabolic, 23\% as sensory and 41\% as metabolic+sensory. We build classification models on these 338 ears using hearing measurements at 0.5, 1, 2, 3, 4, 6, 8 kHz as predictors of the phenotypes.

\begin{figure}[h!]
	\centering
	\includegraphics[width=0.6\linewidth]{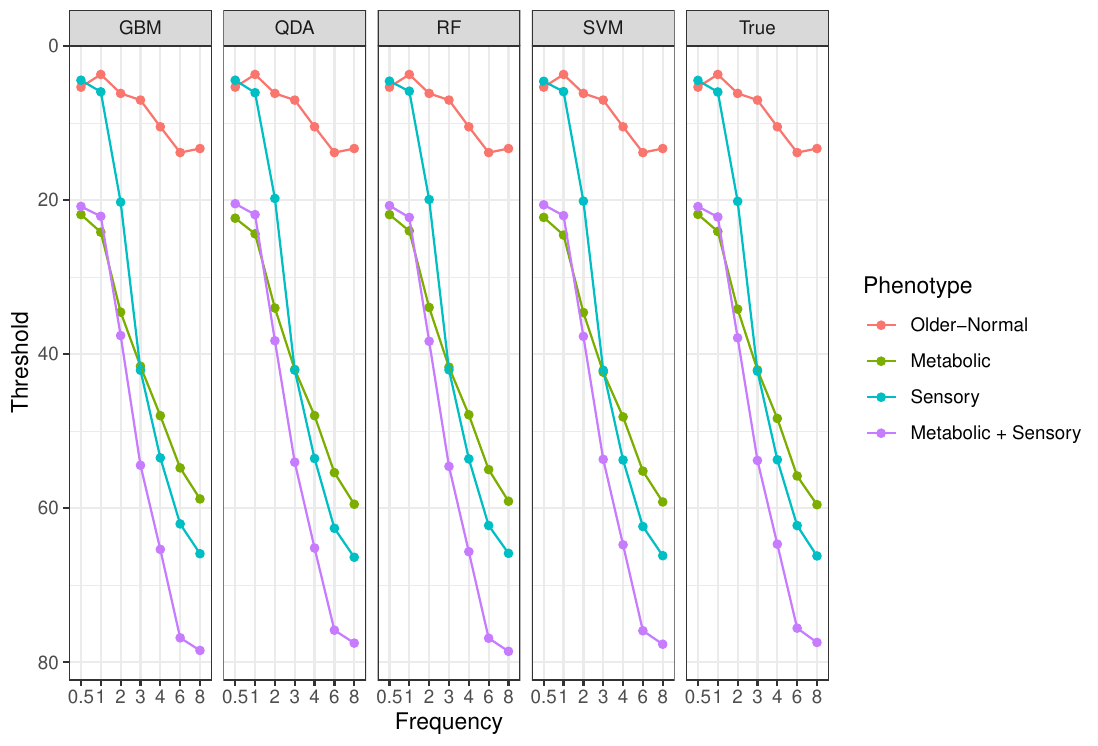}
	\caption{Mean hearing thresholds of the three hearing loss phenotypes and older-normal in MUSC; True: based on the labeled phenotype category in the training dataset; GBM, QDA, RF, SVM: based on the predicted phenotype categories using the corresponding classification approach.}
	\label{train_curve}
\end{figure}

We consider four classification models: Quadratic Discriminant Analysis (QDA), Random Forest (RF), Support Vector Machine (SVM) and Gradient Boosting Machine (GBM). A five-fold cross-validation is applied to tune the hyper-parameters and obtain the prediction accuracy of the models, where QDA and SVM have the highest prediction accuracy: 90.78\% and 90.68\%, respectively, while RF and GBM have slightly worse prediction accuracy: 88.44\% and 86.15\%, respectively. Figure \ref{train_curve} shows the mean hearing thresholds over sound frequencies by the labeled phenotype category in the MUSC training dataset, as well as the mean curves by the predicted phenotype categories based on the four classification approaches, respectively. The models are then applied to make predictions of the phenotypes in AAA, and Supplementary figure 1 shows the mean curves of the hearing thresholds by predicted phenotypes in AAA. 

In addition, we propose to combine the predictions from QDA, RF, SVM and GBM using the stacking regression approach, where the combination weights are obtained by regressing the true labels of ears against the predicted labels from the four classification models. This ensemble approach has the benefit of reducing prediction variation and obtaining more accurate predictions, and we refer readers to \cite{sagi2018ensemble} and \cite{patil2018training} for technical details. In practice, to avoid overfitting, we split the MUSC dataset into two parts. One part is used to train the classification models, and the second part is used to perform the stacking regression to obtain the linear combination weights.

In the main study, the follow-up began in 1991 when the information of study participants' characteristics was mostly available. We explore the associations between tinnitus (yes/no) and metabolic, sensory and metabolic+sensory hearing loss, and we adjust for baseline age, obesity (yes/no), hypertension (yes/no), smoking status (ever/never), and physical activity in the regression \citep{curhan2018adherence}. Tables \ref{real_weighted} and \ref{real_impute} show the cause-specific hazard ratios of tinnitus on metabolic, sensory and metabolic+sensory hearing loss following our proposed weighted and imputed estimating equations approaches, respectively. The point estimates of the hazard ratios from both approaches are not substantially different. We observe significant association between tinnitus and metabolic+sensory hearing loss, while the associations between tinnitus and metabolic or sensory hearing loss are not substantially different.

\begin{table}[h!]
	\caption{Point estimates and 95\% confidence intervals of cause-specific hazard ratio of tinnitus on metabolic, sensory and metabolic+sensory hearing loss. The weighted estimating equations approach is used to fit the model. }
	\begin{tabular}{cccc}
		\hline\hline
		Classification model	& Tinnitus - Metabolic & Tinnitus - Sensory & Tinnitus - Metabolic+Sensory  \\
		\hline
		QDA&1.95 (0.85, 4.45)&1.63 (0.55, 4.89)&3.41 (1.80, 6.54)\\
		Random forest &2.13 (1.08, 4.22)&1.98 (0.72, 5.48)&4.07 (1.95, 8.49)\\
		GBM&2.12 (0.95, 4.75)&2.46 (0.80, 7.58)&4.96 (2.09, 11.78)\\
		SVM&2.10 (1.07, 4.12)&2.28 (0.79, 6.56)&3.99 (1.91, 8.34)\\
		Stacking ensemble&1.75 (0.69, 4.56)&2.88 (0.99, 8.38)&3.75 (1.59, 8.84)\\
		\hline\hline
	\end{tabular}
\label{real_weighted}
\end{table}

\begin{table}[h!]
	\caption{Point estimates and 95\% confidence intervals of cause-specific hazard ratio of tinnitus on metabolic, sensory and metabolic+sensory hearing loss. The imputed estimating equations approach is used to fit the model.}
	\begin{tabular}{cccc}
		\hline\hline
		Classification model	& Tinnitus - Metabolic & Tinnitus - Sensory & Tinnitus - Metabolic+Sensory   \\
		\hline
		QDA&1.71 (0.75, 3.90)&1.28 (0.39, 4.19)&4.08 (2.11, 7.89)\\
		Random forest &1.82 (0.93, 3.59)& 1.61 (0.52, 4.91)& 6.14 (2.92, 12.92)\\
		GBM&2.09 (1.10, 3.96)&1.66 (0.56, 4.91)& 5.00 (2.37, 10.57)\\
		SVM&1.78 (0.90, 3.53)& 2.46 (0.91, 6.63)& 4.82 (2.34, 9.92)\\
		Stacking ensemble&1.25 (0.43, 3.62)&2.25 (0.73, 6.98)&4.04 (1.88, 8.68)\\
		\hline\hline
	\end{tabular}
\label{real_impute}
\end{table}

\section{Discussion}

In this paper, we propose methods to analyze clustered data with unknown event types under the competing risks setting. We propose to use the cause-specific proportional hazards frailty model, where random effects are included in the regression to incorporate possible correlation of the outcome data. To account for the unknown event types, we propose a weighted estimating equations approach, where we replace the event indicators with their corresponding conditional probabilities of occurrence of different event types. The probabilities can be obtained by fitting classification models in a training dataset where the outcome event types are available. In addition, we also propose an imputed estimating equations approach, where we impute the missing event types directly using the predicted event types from the classification models. We also provide analytical variances for the proposed regression coefficient estimates using the robust sandwich variance approach and the inverse of the Hessian matrix of the penalized partial likelihood. It's worth noting that the parameters in the weighted estimating equations can be obtained using existing \texttt{R} package \texttt{coxme} with additional data duplication and manipulation. In the supplementary material, we create a toy example with two competing risks and two correlated units within each cluster to demonstrate the usage of the \texttt{coxme} package to fit our proposed model.


One limitation of our methods is the assumption of no measurement error in the event time. However, this assumption may be violated, especially when the outcomes are based on participants' self reports. When the event times are also subject to measurement error, one potential solution is to consider further reweighting the likelihood contribution from each participant by the probability of events occurring at each potential event time \citep{wu2023outcome}.

\section*{Appendix}
We give the formula for the information matrix. The elements in the information matrix are:
\begin{equation*}
	\begin{split}
		\boldsymbol{\mathcal{I}}_{\boldsymbol{\beta}_k,\boldsymbol{\beta}_k}=&\sum_{i=1}^{N}\sum_{j=1}^{n_i}P(\delta_{ij}=k)\Bigg[  \frac{\sum_{i'=1}^{N}\sum_{j'=1}^{n_i}Y_{i'j'}(t_{ij})\exp\left( \boldsymbol{X}_{i'j'}\boldsymbol{\beta}_k^T+{v}_{i'k} \right)\boldsymbol{X}_{i'j'}\boldsymbol{X}_{i'j'}^T}{\sum_{i'=1}^{N}\sum_{j'=1}^{n_i}Y_{i'j'}(t_{ij})\exp\left( \boldsymbol{X}_{i'j'}\boldsymbol{\beta}_k^T+{v}_{i'k} \right)} \\
		&-\left( \frac{\sum_{i'=1}^{N}\sum_{j'=1}^{n_i}Y_{i'j'}(t_{ij})\exp\left( \boldsymbol{X}_{i'j'}\boldsymbol{\beta}_k^T+{v}_{i'k} \right)\boldsymbol{X}_{i'j'}}{\sum_{i'=1}^{N}\sum_{j'=1}^{n_i}Y_{i'j'}(t_{ij})\exp\left( \boldsymbol{X}_{i'j'}\boldsymbol{\beta}_k^T+{v}_{i'k} \right)}\right)^{\otimes2}\Bigg],
	\end{split}
\end{equation*}

\begin{equation*}
	\begin{split} \boldsymbol{\mathcal{I}}_{\boldsymbol{v},\boldsymbol{v}}=&\sum_{i=1}^{N}\sum_{j=1}^{n_i}\sum_{k=1}^{K}P(\delta_{ij}=k)\Bigg[  \frac{\sum_{i'=1}^{N}\sum_{j'=1}^{n_i}Y_{i'j'}(t_{ij})\exp\left( \boldsymbol{X}_{i'j'}\boldsymbol{\beta}_k^T+{v}_{i'k} \right)\boldsymbol{R}_{i'k}\boldsymbol{R}_{i'k}^T}{\sum_{i'=1}^{N}\sum_{j'=1}^{n_i}Y_{i'j'}(t_{ij})\exp\left( \boldsymbol{X}_{i'j'}\boldsymbol{\beta}_k^T+{v}_{i'k} \right)} \\
		&-\left( \frac{\sum_{i'=1}^{N}\sum_{j'=1}^{n_i}Y_{i'j'}(t_{ij})\exp\left( \boldsymbol{X}_{i'j'}\boldsymbol{\beta}_k^T+{v}_{i'k} \right)\boldsymbol{R}_{i'k}}{\sum_{i'=1}^{N}\sum_{j'=1}^{n_i}Y_{i'j'}(t_{ij})\exp\left( \boldsymbol{X}_{i'j'}\boldsymbol{\beta}_k^T+{v}_{i'k} \right)}\right)^{\otimes2}\Bigg],\\
	\end{split}
\end{equation*}

\begin{equation*}
	\footnotesize
	\begin{split}
		\boldsymbol{\mathcal{I}}_{\boldsymbol{\beta}_k,\boldsymbol{v}}=&\sum_{i=1}^{N}\sum_{j=1}^{n_i}P(\delta_{ij}=k)\Bigg[  \frac{\sum_{i'=1}^{N}\sum_{j'=1}^{n_i}Y_{i'j'}(t_{ij})\exp\left( \boldsymbol{X}_{i'j'}\boldsymbol{\beta}_k^T+{v}_{i'k} \right)\boldsymbol{X}_{i'j'}\boldsymbol{R}_{i'k}^T}{\sum_{i'=1}^{N}\sum_{j'=1}^{n_i}Y_{i'j'}(t_{ij})\exp\left( \boldsymbol{X}_{i'j'}\boldsymbol{\beta}_k^T+{v}_{i'k} \right)} \\
		&-\left( \frac{\sum_{i'=1}^{N}\sum_{j'=1}^{n_i}Y_{i'j'}(t_{ij})\exp\left( \boldsymbol{X}_{i'j'}\boldsymbol{\beta}_k^T+{v}_{i'k} \right)\boldsymbol{X}_{i'j'}}{\sum_{i'=1}^{N}\sum_{j'=1}^{n_i}Y_{i'j'}(t_{ij})\exp\left( \boldsymbol{X}_{i'j'}\boldsymbol{\beta}_k^T+{v}_{i'k} \right)}\right)  \left( \frac{\sum_{i'=1}^{N}\sum_{j'=1}^{n_i}Y_{i'j'}(t_{ij})\exp\left( \boldsymbol{X}_{i'j'}\boldsymbol{\beta}_k^T+{v}_{i'k} \right)\boldsymbol{R}_{i'k}}{\sum_{i'=1}^{N}\sum_{j'=1}^{n_i}Y_{i'j'}(t_{ij})\exp\left( \boldsymbol{X}_{i'j'}\boldsymbol{\beta}_k^T+{v}_{i'k} \right)}\right)^T\Bigg].
	\end{split}
\end{equation*}
\bibliography{ref}

\begin{thebibliography}{20}
\providecommand{\natexlab}[1]{#1}
\providecommand{\url}[1]{\texttt{#1}}
\expandafter\ifx\csname urlstyle\endcsname\relax
  \providecommand{\doi}[1]{doi: #1}\else
  \providecommand{\doi}{doi: \begingroup \urlstyle{rm}\Url}\fi

\bibitem[Craiu and Duchesne(2004)]{craiu2004inference}
R.~V. Craiu and T.~Duchesne.
\newblock Inference based on the em algorithm for the competing risks model
  with masked causes of failure.
\newblock \emph{Biometrika}, 91\penalty0 (3):\penalty0 543--558, 2004.

\bibitem[Curhan et~al.(2018)Curhan, Wang, Eavey, Stampfer, and
  Curhan]{curhan2018adherence}
S.~G. Curhan, M.~Wang, R.~D. Eavey, M.~J. Stampfer, and G.~C. Curhan.
\newblock Adherence to healthful dietary patterns is associated with lower risk
  of hearing loss in women.
\newblock \emph{The Journal of nutrition}, 148\penalty0 (6):\penalty0 944--951,
  2018.

\bibitem[Curhan et~al.(2021)Curhan, Halpin, Wang, Eavey, and
  Curhan]{curhan2021tinnitus}
S.~G. Curhan, C.~Halpin, M.~Wang, R.~D. Eavey, and G.~C. Curhan.
\newblock Tinnitus and 3-year change in audiometric hearing thresholds.
\newblock \emph{Ear and hearing}, 42\penalty0 (4):\penalty0 886, 2021.

\bibitem[Dharmarajan et~al.(2018)Dharmarajan, Schaubel, and
  Saran]{dharmarajan2018evaluating}
S.~H. Dharmarajan, D.~E. Schaubel, and R.~Saran.
\newblock Evaluating center performance in the competing risks setting:
  Application to outcomes of wait-listed end-stage renal disease patients.
\newblock \emph{Biometrics}, 74\penalty0 (1):\penalty0 289--299, 2018.

\bibitem[Dubno et~al.(2013)Dubno, Eckert, Lee, Matthews, and
  Schmiedt]{dubno2013classifying}
J.~R. Dubno, M.~A. Eckert, F.-S. Lee, L.~J. Matthews, and R.~A. Schmiedt.
\newblock Classifying human audiometric phenotypes of age-related hearing loss
  from animal models.
\newblock \emph{Journal of the Association for Research in Otolaryngology},
  14:\penalty0 687--701, 2013.

\bibitem[Goetghebeur and Ryan(1995)]{goetghebeur1995analysis}
E.~Goetghebeur and L.~Ryan.
\newblock Analysis of competing risks survival data when some failure types are
  missing.
\newblock \emph{Biometrika}, 82\penalty0 (4):\penalty0 821--833, 1995.

\bibitem[Gray(1992)]{gray1992flexible}
R.~J. Gray.
\newblock Flexible methods for analyzing survival data using splines, with
  applications to breast cancer prognosis.
\newblock \emph{Journal of the American Statistical Association}, 87\penalty0
  (420):\penalty0 942--951, 1992.

\bibitem[Gupta et~al.(2019)Gupta, Eavey, Wang, Curhan, and
  Curhan]{gupta2019type}
S.~Gupta, R.~D. Eavey, M.~Wang, S.~G. Curhan, and G.~C. Curhan.
\newblock Type 2 diabetes and the risk of incident hearing loss.
\newblock \emph{Diabetologia}, 62:\penalty0 281--285, 2019.

\bibitem[Hougaard(1995)]{hougaard1995frailty}
P.~Hougaard.
\newblock Frailty models for survival data.
\newblock \emph{Lifetime data analysis}, 1:\penalty0 255--273, 1995.

\bibitem[Hyun et~al.(2012)Hyun, Lee, and Sun]{hyun2012proportional}
S.~Hyun, J.~Lee, and Y.~Sun.
\newblock Proportional hazards model for competing risks data with missing
  cause of failure.
\newblock \emph{Journal of statistical planning and inference}, 142\penalty0
  (7):\penalty0 1767--1779, 2012.

\bibitem[Lin et~al.(2020)Lin, Wang, Stankovic, Eavey, McKenna, Curhan, and
  Curhan]{lin2020cigarette}
B.~M. Lin, M.~Wang, K.~M. Stankovic, R.~Eavey, M.~J. McKenna, G.~C. Curhan, and
  S.~G. Curhan.
\newblock Cigarette smoking, smoking cessation, and risk of hearing loss in
  women.
\newblock \emph{The American journal of medicine}, 133\penalty0 (10):\penalty0
  1180--1186, 2020.

\bibitem[Lok(2021)]{lok2021estimating}
J.~J. Lok.
\newblock How estimating nuisance parameters can reduce the variance (with
  consistent variance estimation).
\newblock \emph{arXiv preprint arXiv:2109.02690}, 2021.

\bibitem[Lu and Tsiatis(2001)]{lu2001multiple}
K.~Lu and A.~A. Tsiatis.
\newblock Multiple imputation methods for estimating regression coefficients in
  the competing risks model with missing cause of failure.
\newblock \emph{Biometrics}, 57\penalty0 (4):\penalty0 1191--1197, 2001.

\bibitem[Patil and Parmigiani(2018)]{patil2018training}
P.~Patil and G.~Parmigiani.
\newblock Training replicable predictors in multiple studies.
\newblock \emph{Proceedings of the National Academy of Sciences}, 115\penalty0
  (11):\penalty0 2578--2583, 2018.

\bibitem[Prentice et~al.(1978)Prentice, Kalbfleisch, Peterson~Jr, Flournoy,
  Farewell, and Breslow]{prentice1978analysis}
R.~L. Prentice, J.~D. Kalbfleisch, A.~V. Peterson~Jr, N.~Flournoy, V.~T.
  Farewell, and N.~E. Breslow.
\newblock The analysis of failure times in the presence of competing risks.
\newblock \emph{Biometrics}, pages 541--554, 1978.

\bibitem[Ripatti and Palmgren(2000)]{ripatti2000estimation}
S.~Ripatti and J.~Palmgren.
\newblock Estimation of multivariate frailty models using penalized partial
  likelihood.
\newblock \emph{Biometrics}, 56\penalty0 (4):\penalty0 1016--1022, 2000.

\bibitem[Sagi and Rokach(2018)]{sagi2018ensemble}
O.~Sagi and L.~Rokach.
\newblock Ensemble learning: A survey.
\newblock \emph{Wiley Interdisciplinary Reviews: Data Mining and Knowledge
  Discovery}, 8\penalty0 (4):\penalty0 e1249, 2018.

\bibitem[Therneau et~al.(2003)Therneau, Grambsch, and
  Pankratz]{therneau2003penalized}
T.~M. Therneau, P.~M. Grambsch, and V.~S. Pankratz.
\newblock Penalized survival models and frailty.
\newblock \emph{Journal of computational and graphical statistics}, 12\penalty0
  (1):\penalty0 156--175, 2003.

\bibitem[Wang et~al.(2020)Wang, He, and Schaubel]{wang2020penalized}
L.~Wang, K.~He, and D.~E. Schaubel.
\newblock Penalized survival models for the analysis of alternating recurrent
  event data.
\newblock \emph{Biometrics}, 76\penalty0 (2):\penalty0 448--459, 2020.

\bibitem[Wu and Wang(2023)]{wu2023outcome}
Y.~Wu and M.~Wang.
\newblock Outcome measurement error correction for survival analyses with
  multiple failure types: application to hearing loss studies.
\newblock \emph{arXiv preprint arXiv:2306.10568}, 2023.

\end{thebibliography}

\end{document}